\documentclass[aps,preprint]{revtex4}%
\usepackage{amsfonts}
\usepackage{amsmath}
\usepackage{amssymb}
\usepackage{graphicx}%
\begin{document}
\title[Short title for running header]{The Zinc Doping Induced STM Resonance as a Zero Mode}
\author{Tao Li}
\affiliation{Center for Advanced Study, Tsinghua University, Beijing 100084, P.R.China}

\begin{abstract}
We show that the Zinc doping induced STM resonance in high-Tc cuprates is a
zero mode - a d-wave monomer confined in the opposite sublattice of the Zinc
site. We propose that the resonance come from the superconducting peak around
$\left(  \pi,0\right)  $ as observed in ARPES experiments. It is predicted
that the resonance should vanish in single layer $BSCCO_{2201}$ or in the
vortex core of underdoped cuprate where the ARPES peak is absent. We argue
that the STM result imply the ARPES peak around $\left(  \pi,0\right)  $ is
the only coherent feature in the whole Brillouin zone. 

\end{abstract}
\volumeyear{year}
\volumenumber{number}
\issuenumber{number}
\eid{identifier}
\startpage{1}
\endpage{10}
\maketitle

\bigskip Zinc doping is an important probe of the high temperature
superconductors for two reasons. On the one hand, it is a clean probe. The
closed shell $Zn^{2+}$ ion substitute for the $Cu^{2+}$ ion in the $CuO_{2}$
plane and act almost as a vacancy of the lattice and is thus almost parameter
free. On the other hand, experiments find dramatic effects of zinc doping in
almost all important detecting channels such as transport, spin dynamics and
single particle spectrum.

Recently, STM experiments find a sharp resonance near zero energy around the
Zinc sites in $BSCCO_{2212}$\cite{1}. Many theoretical efforts have been
devoted to extract information from this remarkable
phenomena\cite{2,3,4,5,6,7}. Roughly speaking, there are two class of
theories. The first kind of theory view the resonance as the result of
potential scattering of d-wave BCS quasiparticle from the Zinc site. The
second kind of theory view the resonance as a result of Kondo effect of the
nodal quasiparticle with the presumed Zinc-doping induced local moment.
Although experiment result do show some similarity with the prediction of
these theories, no theory is truly satisfactory. The following characteristics
of the resonance are especially difficult to understand. (1)the resonance is
extremely sharp in energy and extremely close to zero energy. (2)the spectral
weight of the resonance seems to be transfered totally from the so called
coherence peak in the density of state, with other part of the spectrum
essentially unchanged(except for a nearly constant downward shift of spectral
weight in the particle side). (3)the resonance extend mainly in the antinodal
direction rather than the nodal direction in real space. (4)The resonance
exist only on the sublattice that the doped Zinc ion resides. Since the
observed STM spectrum at any site is actually the sum of spectrums from its
four nearest-neighbouring sites(which are on the opposite sublattice of the
given site) due to the filtering of the $BiO$ layer intervening the STM needle
and the $CuO_{2}$ plane\cite{7}, this indicate that the resonance is confined
on the opposite sub-lattice of Zinc site. In the first kind of theory, the
resonance mode is well defined only when the system is particle-hole symmetric
and the scattering potential is infinite(the resonance energy is zero in this
case). However, this symmetry exist only in the half-filled case and is broken
in doped system. Many body effect make the situation even worse. For example,
STM spectrum shows a large downward slope in the density of state near Fermi
level which is attributed to strong correlation effect by some author\cite{8}.
The spectral weight transfer pattern and the extending direction of the
resonance in real space predicted by these kind of theories are also
problematic. Calculation shows the spectrum change on all energy scale upon
Zinc doping even in the particle-hole symmetric case. Correspondingly, the
resonance mode always change its extending direction into the nodal direction
at large distance since the Fermi velocity is maximum in that direction. For
the Kondo-like theories, the most severe problem is that the nature of the
presumed local moment and its coupling to the\ nodal quasiparticle are not
clear. At the same time, this mechanism suffers from the same problem as the
first kind of theories as regard to the spectral weight transfer pattern and
extending direction of the mode in real space since nodal quasiparticle play
the central role in this framework. There is also no way to understand the
sublattice selection rule in this framework. We note that the Kondo screening
vanishes in the particle-hole symmetric case and thus the two kind of
interpretations of the resonance are in fact orthogonal\cite{5}.

We take the doped $Zn^{2+}$ ion as a vacancy of the lattice. We show that the
Zinc doping induced single particle resonance is a zero mode confined in the
opposite sublattice of the Zinc site. This zero mode is in fact the d-wave
partner of the Zinc site whose pairing with the Zinc site is made ineffective
by the large potential difference between the two participants of the pairing.
Since this is a d-wave state, it spectral weight comes mainly from the
antinodal region of the Brillouin Zone and has no contribution from
quasiparticle exact along the nodal direction. Guided by this picture, we
propose that this resonance comes from the so called superconducting peak near
$\left(  \pi,0\right)  $\bigskip\ observed in ARPES experiments\cite{9}. This
proposal has the virtue that the particle-hole symmetry is easily satisfied
since the ARPES peak exist only in a small region around $\left(
\pi,0\right)  $ in which there is almost no dispersion. It also build in
naturally the correct spectral weight transfer pattern and correct extending
direction of the resonance in real space. This proposition has many
interesting predictions. For example, it predicts that the resonance should
vanish in vortex core with a pseudogap and in single layer $BSCCO_{2201}%
$\cite{9} where no ARPES peak is observed. This picture also indicated that
the ARPES peak around $\left(  \pi,0\right)  $ is the only coherent spectral
feature in the whole Brillouin Zone.

The doped Zinc ion break the translational symmetry of the lattice. However,
the point group symmetry with respect to the Zinc site is preserved(here we
assume a single Zinc ion for simplicity). Hence we assume that the Zinc doped
system preserve the d-wave symmetry of superconducting pairing under the point
group operation. With the pairing being d-wave, a s-wave particle can only
pair with a d-wave particle and vice versa. Hence the particle that pair with
the Zinc site must be d-wave. Since the kinetic energy of the Hamiltonian is
symmetric under the point group operation, such a d-wave particle has no
chance to feel the strong s-wave repulsive potential on the Zinc site. Hence
the zinc site and its d-wave pairing partner feel a large potential difference
due to the symmetry mismatch of pairing and scattering potential. This large
potential difference make the pairing ineffective and we simply get a level
pushed to large energy and the other essentially unchanged from the its bare
energy. In other words, the pairing is broken and we are left with a d-wave
particle missing its pairing partner(the missing particle is just the Zinc
site). This d-wave monomer, is nothing but the observed resonance in
STM\ spectrum. This picture of the STM resonance as a d-wave monomer naturally
explain the fact that the spectral weight of the resonance comes mainly from
the antinodal region and that the resonance extend mainly in the antinodal
direction. This argument is, however, oversimplified in that we have totally
neglected the dispersion caused by the kinetic energy and the pairing
potential. With these dispersion present, we generally only get a edge state -
a virtual bound state with finite life time and nonzero energy, rather than a
well defined mode at zero energy. To show this more directly, we make a
T-matrix calculation of the impurity effect in a d-wave BCS state. Here we
model the Zinc impurity as a repulsive potential scatter with strength $U$ at
the origin. The real space Green's function is given by%

\[
g_{i,j}=g_{i,j}^{0}+g_{i,0}^{0}U\tau^{3}[I-g_{0,0}^{0}U\tau^{3}]^{-1}%
g_{0,j}^{0}%
\]

here $g_{i,j}^{0}$ is the Green's function of the pure system in the Nambu's
notation and is given by the Fourier transform of the momentum space Green's function,%

\[
g_{i,j}^{0}=\frac{1}{\left(  2\pi\right)  ^{2}}\int dk^{2}\frac{e^{ik(R_{i}%
-R_{j})}}{\omega+i\delta-\xi_{k}\tau^{3}+\Delta_{k}\tau^{1}}%
\]
$\ \tau^{1},\tau^{3}$ are the Pauli matrix appearing in the Nambu notation,
$I$ is the identity matrix, $\xi_{k}$ and $\Delta_{k}$ are bare dispersion and
pairing order parameter. Here we take $\xi_{k}=-2(\cos(k_{x})+\cos(k_{y}%
))-\mu$ , $\Delta_{k}=2(\cos(k_{x})-\cos(k_{y}))$. In this case, the
particle-hole symmetry is explicitly broken by the finite chemical potential
at nonzero doping(in general the particle-hole symmetry can also be broken by
introducing longer range hopping term that make the lattice non-bipartite).
The additional spectral weight caused by Zinc doping come from the pole of the
$[I-g_{0,0}^{0}U\tau^{3}]^{-1}$. The pole equation has a zero energy solution
when the off-diagonal matrix element of $g_{0,0}^{0}$ vanish and the system is
particle-hole symmetric(which also imply that $U$ is infinite). The first
condition correspond to the physical requirement that the pairing partner(not
necessarily d-wave) of the Zinc site should have zero amplitude at the origin
and so can not feel the strong repulsive potential. The second condition,
namely the requirement of particle-hole symmetry, in fact imply that the
system is bipartite. To show this, we take a general pairing Hamiltonian on
the lattice(not necessarily translational invariant)

\bigskip%
\begin{align*}
H  &  =%
{\displaystyle\sum\limits_{i,,j,\sigma}}
t_{i,j}c_{i,\sigma}^{\dagger}c_{j,\sigma}+%
{\displaystyle\sum\limits_{i,j}}
(\Delta_{i,j}c_{i,\uparrow}^{\dagger}c_{j,\downarrow}^{\dagger}+h.c.)+%
{\displaystyle\sum\limits_{i,\sigma}}
\mu_{i}c_{i,\sigma}^{\dagger}c_{i,\sigma}\\
&  =H_{t}+H_{\Delta}+H_{\mu}%
\end{align*}

\bigskip and examine how it transform under the following particle-hole transformation%

\begin{align*}
c_{i,\uparrow}^{\dagger}  &  \longrightarrow c_{i,\downarrow}\\
c_{i,\downarrow}^{\dagger}  &  \longrightarrow c_{i,\uparrow}%
\end{align*}

. Under this transformation, the three part of $H$ transform as follow(here we
restrict ourself to the case of singlet pairing with a real $\Delta_{i,j}$ for
simplicity), \ 

\bigskip%
\begin{align*}
H_{t}  &  \longrightarrow-H_{t}\\
H_{\Delta}  &  \longrightarrow H_{\Delta}\\
H_{\mu}  &  \longrightarrow-H_{\mu}+2%
{\displaystyle\sum\limits_{i}}
\mu_{i}%
\end{align*}

For a particle-hole symmetric system, the transformed Hamiltonian should be
gauge equivalent to the original one. This requirement has three implications.
First, $H_{\mu}$ should vanish since it is gauge invariant. Second, the
lattice should be bipartite in the sense of $H_{t}$ since otherwise we can
always construct a odd-bond loop in which the enclosed flux differ by $\pi$
before and after the transformation$.$Third, the pairing term should be
inter-sublattice only or intra-sublattice only as can be shown by a similar
argument provided that the lattice is connected in the sense of $H_{t}$(that
is, any two sites of the lattice can be connected by a path made up of bonds
with nonzero $t_{i,j}$). For the high temperature superconductors that we are
interesting in, the pairing term should be inter-sublattice only for the
system to be particle-hole symmetric. Hence the whole Hamiltonian is
bipartite(the intra-sublattice pairing case is also bipartite, but in a
different sense). For a bipartite system with a vacancy in one sublattice, it
is easy to show a mode will appear exactly at zero energy which is confined on
the opposite sublattice of the vacancy. The proof of this statement can be
briefed as follow. Let $\psi$ be an eigen mode of the system(i.e. $\left[
\psi,H\right]  =E\psi$). We can always write $\psi$ as $\psi_{1}+\psi_{2}$, in
which $\psi_{1\text{ }}$and $\psi_{2}$ denote the components of $\psi$ in the
two sublattices. Since the Hamiltonian is bipartite, we have $\left[  \psi
_{1},H\right]  =E\psi_{2}$ and $\left[  \psi_{2},H\right]  =E\psi_{1}$. Hence
for nonzero $E,$ both $\psi_{1}$ and $\psi_{2}$ are nonzero and $\psi_{1}%
-\psi_{2}$ is a eigen mode of energy $-E$. At the same time, since $\left[
[\psi_{1},H],H\right]  =[\psi_{1},H^{2}]=E^{2}\psi_{1}$ and $\left[  [\psi
_{2},H],H\right]  =[\psi_{2},H^{2}]=E^{2}\psi_{2}$, $\psi_{1}$ and $\psi_{2}$
are eigen modes of the Hermite operator $H^{2}$ and thus form complete
orthogonal basis in the Hilbert space of the two sublattices respectively(for
degenerate $E,$ there exist an orthogonal procedure that make each components
orthogonal in their respective subspace, the property $\left\{  \psi_{1}%
,\psi_{1}^{\dagger}\right\}  =\left\{  \psi_{2},\psi_{2}^{\dagger}\right\}  $
which follows from the fact that $\left\{  \psi_{1}+\psi_{2},\psi_{1}%
^{\dagger}-\psi_{2}^{\dagger}\right\}  =0$ is useful in proving this). Thus,
nonzero energy eigen mode always appear in pairs and exploit one state from
each of the two sublattice. Since the dimension of the Hilbert space of the
two sublattice differ by one, there should always be a state on the opposite
sublattice of the vacancy that is left unpaired and have zero energy. This
mode is in fact a zero mode in general context which exists for topological
reason. In our case, the dimension of the Hilbert space of the two sublattices
differ by one due to the vacancy. This is analogues\ to the topological
constraint near a domain wall in the 1d Peieris system in which case a zero
mode exist\cite{10}.

The picture of the STM resonance as a d-wave monomer do capture the main
characteristics of the experimental result. However, it is still not a
interpretation of the experiment until we understand the following points.
First, where is the required particle-hole symmetry? As we have mentioned,
this symmetry is badly violated in the high temperature superconductors.
Second, the d-wave nature of the mode alone is not enough to filter
contribution from the nodal region totally. These remaining contribution from
the nodal region will dominate the long distance behavior of the mode since
the Fermi velocity is maximal in that region and we expect the mode to extend
mainly in the nodal direction at sufficient large distance. 

Fortunately, our picture also tell us where to find the way out. Since the
spectral weight of the mode is dominated by contribution from the antinodal
region, we should focus our attention at this region. Experimentally, ARPES
show a characteristic 'peak-dip-hump' structure in the antinodal region below
the superconducting transition temperature\cite{9}. The peak intensity
increase with decreasing temperature in a way similar to that of superfluid
density. We propose that the Zinc doping induced STM resonance comes from this
ARPES peak. This proposition has the virtue that the all important
particle-hole symmetry is easily satisfied since the ARPES peak exist only in
a small region around $\left(  \pi,0\right)  $ in which there is almost no
dispersion. At the same time, this mechanism also naturally built in the
correct spectral weight transfer pattern and the correct extending direction
of the mode in real space since it is free from the concomitant of nodal
contribution.\bigskip\ Since the ARPES peak is dispersionless, we model it as
a Fermi surface mode with a d-wave pairing gap of constant amplitude $\Delta$.
In this case, the equation for the pole of $[I-g_{0,0}^{0}U\tau^{3}]^{-1}$
reduce to%

\[
\left[  1-U\alpha\frac{\omega}{\omega^{2}-\Delta^{2}}\right]  \left[
1+U\alpha\frac{\omega}{\omega^{2}-\Delta^{2}}\right]  =0
\]

here the parameter $\alpha$ take care of the effect of both wave function
renormalization and the vertex renormalization. The solution of this equation
is given by%

\[
E=\frac{\pm U\alpha\pm\sqrt{(U\alpha)^{2}+4\Delta^{2}}}{2}%
\]

for large $U$, the solution can be approximated by $E_{1}=\pm U\alpha$ and
$E_{2}=\pm\frac{\Delta^{2}}{U\alpha}$. The second solution is the zero mode we
discussed in this paper. In principle, we can calculate the real space
distribution of the mode if we know the momentum distribution of the ARPES
peak in the Brillouin Zone. Roughly speaking, the spatial range of the mode in
real space is inversely related to the momentum range of the ARPES peak in the
Brillouin Zone. The extending direction of the mode in real space is
determined by the anisotropy of the momentum region in which the ARPES peak
exist. As a simple model, we take the momentum region of the ARPES peak as the
rectangular area around $\left(  \pi,0\right)  $ shown in Figure 1. This
assignment is reasonable since the Fermi surface near $\left(  \pi,0\right)  $
is parallel to the antinodal direction\cite{9}. Figure 2 shows the calculated
spectrum on the nearest neighbouring site of Zinc, Figure 3 shows the
calculated real space distribution of local density of state contributed by
the resonance mode. The results are within our theoretical expectation and
consistent with experimental results.

\bigskip Our proposition for the resonance mode has many interesting
predictions. According to our proposition, the STM resonance should vanish
with the ARPES peak. We can check this prediction in single layer
$BSCCO_{2201}$ since ARPES find no 'peak-dip-hump' structure around $\left(
\pi,0\right)  $ in this material. We can also check the prediction in a
magnetic field since it is believed that magnetic field can reduce the
intensity of the ARPES peak. Especially, the ARPES peak will vanish totally in
a pseudogap phase vortex core and we expect the STM resonance to disappear in
the vortex core of a underdoped cuprate.

In our theory, the resonance mode always appear in pairs on both side of bias.
For a repulsive scatter, the main spectral weight is particle-like. However,
the resonance observed in experiment appear at small negative bias. This means
that the resonance mode is occupied by an unpaired electron. This is supported
by the nearly constant downward shift of spectral weight in the particle side
of the spectrum which imply a local increase of electron density. This
conclusion can also be reached by examine the extending direction of the mode
in real space since a hole mode should extend mainly in the nodal direction
due to the complementary nature particle and hole. According to this picture,
we should expect a secondary peak below the main peak which extend mainly in
the nodal direction. We note experiments do show a small peak below the main
peak. It is interesting to check its extending direction in real space.
Mathematically, the ground state of the Zinc-doped system should of the form
$\gamma_{e}^{\dagger}\left\vert paired\right\rangle $. Here $\gamma
_{e}^{\dagger}$ is electron-like excitation. The main STM peak should
correspond to the process $\gamma_{e}^{\dagger}\left\vert paired\right\rangle
\longrightarrow\left\vert paired\right\rangle $. The secondary peak should
correspond to the process $\gamma_{e}^{\dagger}\left\vert paired\right\rangle
\longrightarrow\gamma_{e}^{\dagger}\gamma_{h}^{\dagger}\left\vert
paired\right\rangle $, where $\gamma_{h}^{\dagger}$ is a hole-like excitation.

In our discussion, we have neglect the coupling of the ARPES peak with other
excitations in the Brillouin Zone. This neglect has important implication.
Technically, this coupling can be expressed in terms of the following matrix element%

\[
\left\langle \text{ARPES Peak}\right\vert \hat{U}\left\vert \text{Other
excitations}\right\rangle
\]

where $\hat{U}$ is the potential scattering. The vanishing of this coupling
indicate that the ARPES peak and other excitations in the Brillouin Zone are
orthogonal in nature. If we assume the ARPES peak is a coherent object(as
suggested by its relation with the superfluid density), we must conclude that
all other excitations in Brillouin Zone are incoherent, that is, their Fermi
liquid spectral weight $Z$ should be zero. Especially, the nodal excitation
should be incoherent. This may already be confirmed by ARPES experiment in
both normal and superconducting state\cite{11}.

\bigskip\bigskip The author would like to thank members of the HTS group at
CASTU for discussion.

\begin{center}
\newpage

FIGURES
\end{center}

FIG. 1. Momentum region of the ARPES peak in the Brillouin zone.

\medskip

FIG. 2. Local density of state at the nearest-neighbouring site of Zinc
contributed by the resonance. The thin line shows the contribution from ARPES
peak in pure system.\medskip

FIG. 3. Real space distribution of the resonance. (a)Before $BiO$ layer
filtering. (b)After $BiO$ layer filtering.

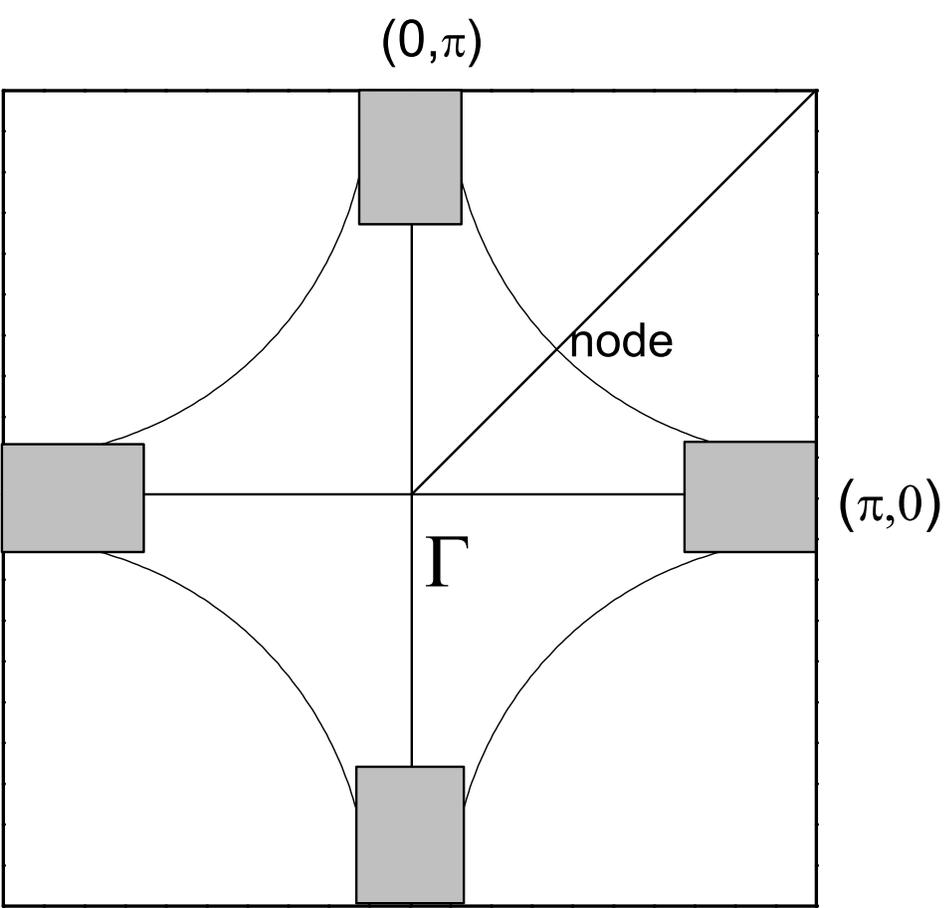
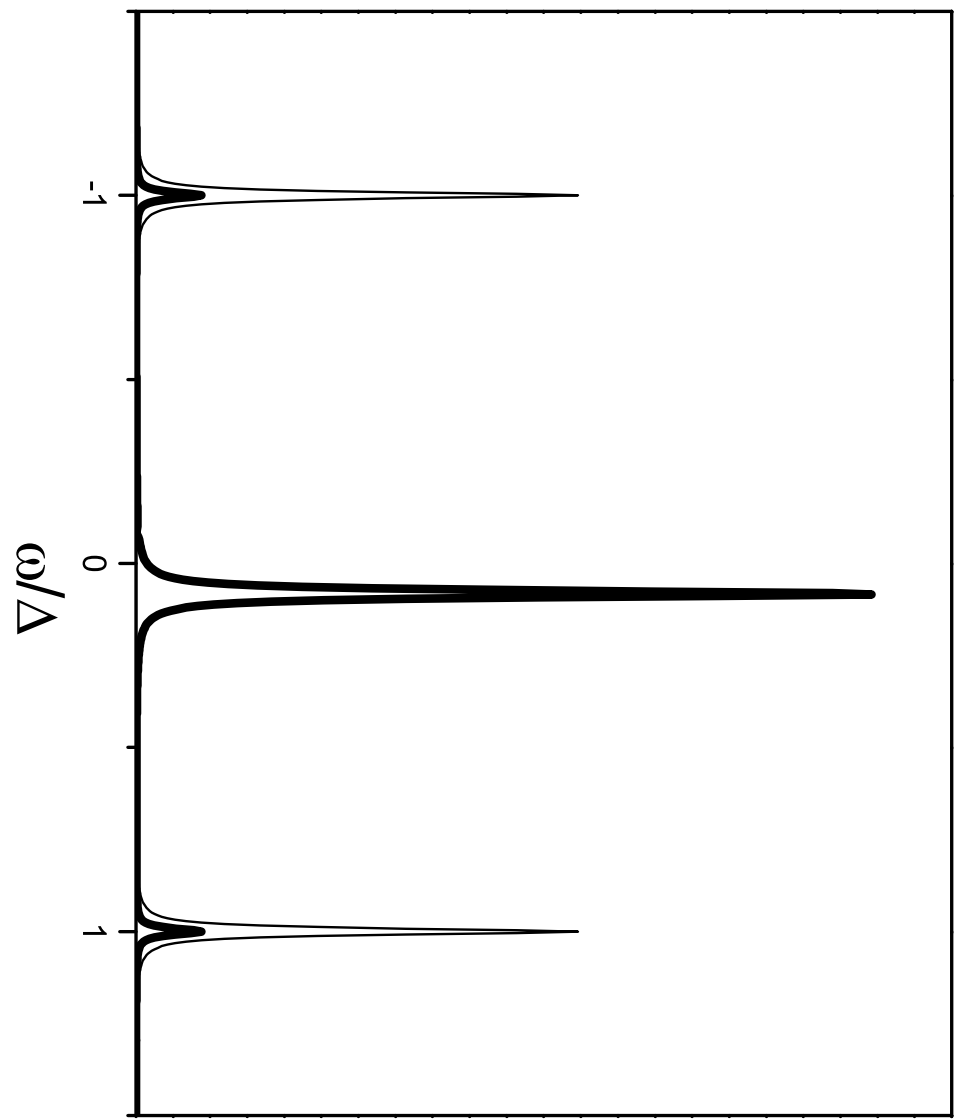
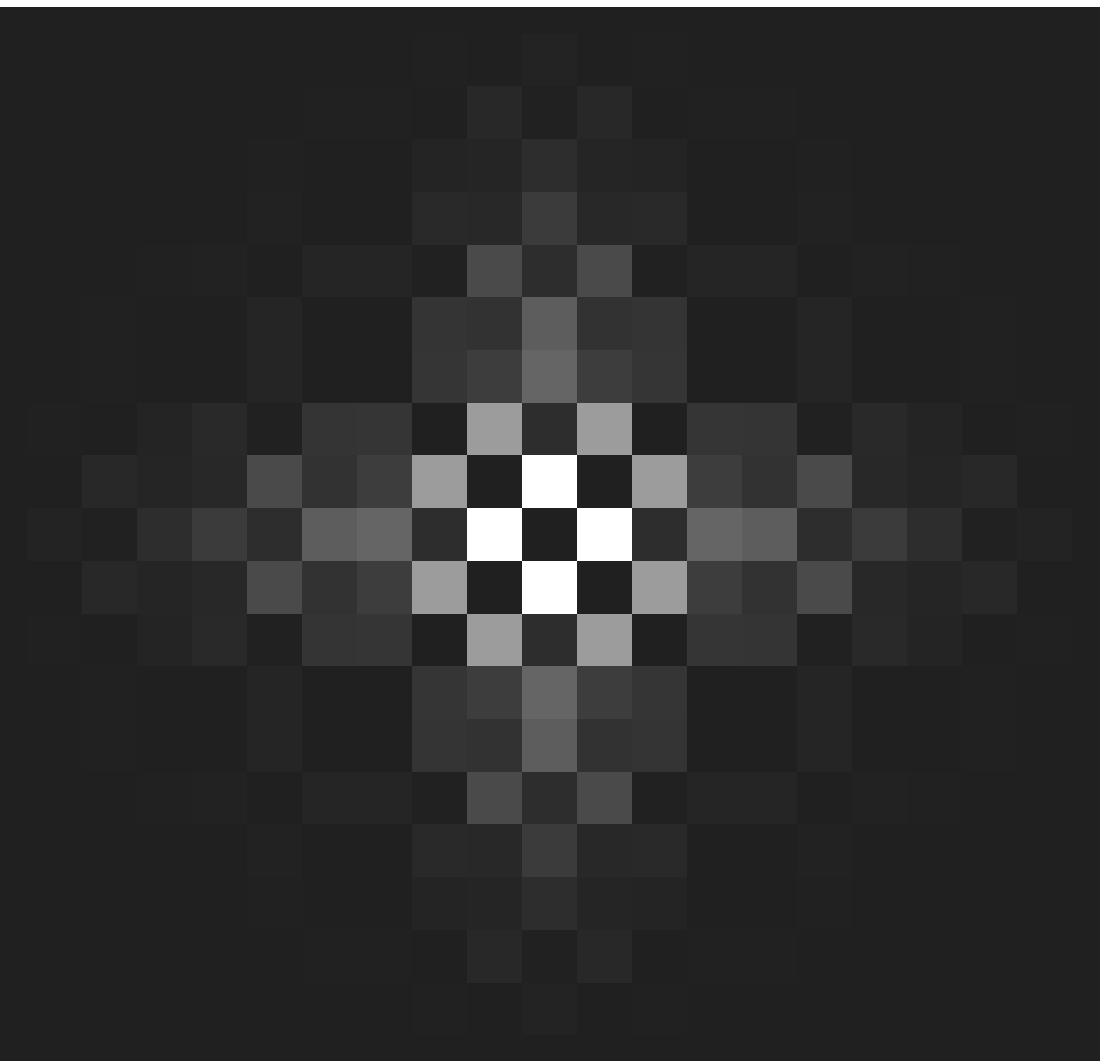
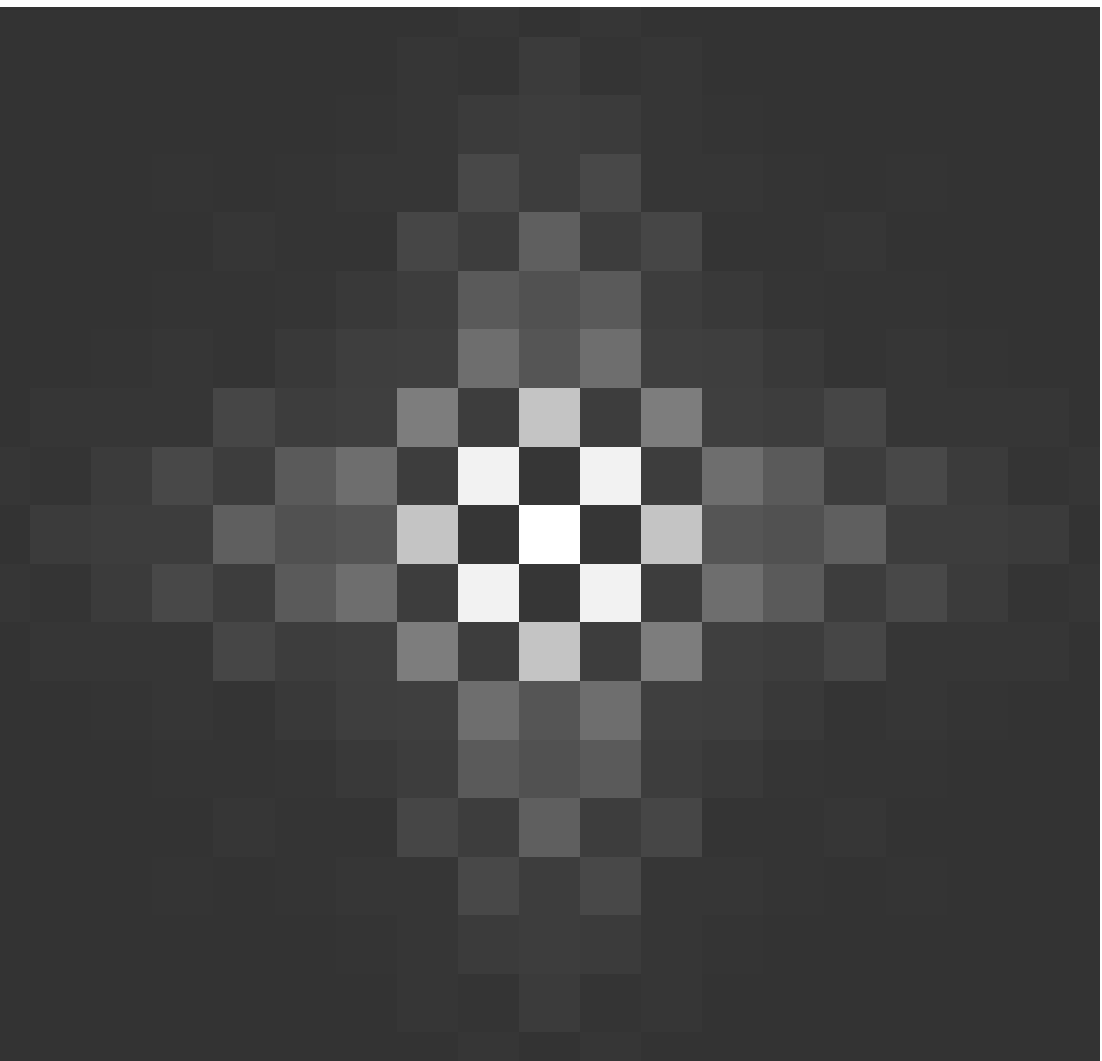
\end{document}